# Thermal Conductivity of CaSiO$_3$ Perovskite at Lower Mantle Conditions

Zhen Zhang[1], Dong-Bo Zhang[2,3], Kotaro Onga[4], Akira Hasegawa[4,5], Kenji Ohta[4], Kei Hirose[6,7] & Renata M. Wentzcovitch[1,8,9]*

[1]Department of Applied Physics and Applied Mathematics, Columbia University, New York, NY 10027, USA.

[2]College of Nuclear Science and Technology, Beijing Normal University, Beijing 100875, People's Republic of China.

[3]Beijing Computational Science Research Center, Beijing 100193, People's Republic of China.

[4]Department of Earth and Planetary Sciences, Tokyo Institute of Technology, Meguro, Tokyo 152-8551, Japan.

[5]National Metrology Institute of Japan, National Institute of Advanced Industrial Science and Technology, Tsukuba, Ibaraki 305-8563, Japan.

[6]Department of Earth and Planetary Science, The University of Tokyo, Bunkyo, Tokyo 113-0033, Japan.

[7]Earth-Life Science Institute, Tokyo Institute of Technology, Meguro, Tokyo 152-8550, Japan.

[8]Department of Earth and Environmental Sciences, Columbia University, New York, NY 10027, USA.

[9]Lamont–Doherty Earth Observatory, Columbia University, Palisades, NY 10964, USA.

*email: rmw2150@columbia.edu

**Thermal conductivity (κ) of mantle minerals is a fundamental property in geodynamic modeling. It controls the style of mantle convection and the time scale of the mantle and core cooling. Cubic $CaSiO_3$ perovskite (CaPv) is the third most abundant mineral in the lower mantle (LM) (7 vol%). However, despite its importance, no theoretical or experimental estimate of CaPv's κ exists. Theoretical investigations of its properties are challenging because of its strong anharmonicity. Experimental measurements at relevant high pressures and temperatures are equally challenging. Here we present *ab initio* results for CaPv's κ obtained using an established phonon quasiparticle approach that can address its strong anharmonicity. We also offer experimental measurements of κ . Predictions and measurements are in good agreement and reveal a surprisingly large κ for cubic CaPv. Despite its relatively low abundance, CaPv's κ might increase the lower mantle κ by approximately 10%, if accounted for. κ of mantle regions enriched in crust material will be more strongly impacted.**

The dominant mode of heat transport between the Earth's core and surface controls the planet's internal dynamics and evolution. The core-mantle boundary (CMB) located at 2,890 km depth is the interface between the molten metallic core and the rocky mantle. At the CMB, where mass transport is impeded, the lower mantle (LM) receives heat released by the core via a conduction mechanism[1]. Knowledge of the thermal conductivity ($\kappa$) of LM minerals is critical for constraining the CMB heat flow[2], which provides a basis for understanding the dynamic state and the thermal history of the mantle and the core, as well as the relative importance of conduction vs. convection above the CMB. The LM extends from 670 to 2,890 kilometers in depth, corresponding to 55 vol% of the Earth's whole interior. Pressures (*P*) and temperatures (*T*) in this region vary between 23 <

$P < 135$ GPa and $2{,}000 < T < 4{,}000$ K, respectively[3,4]. Such extreme conditions introduce considerable challenges for both direct measurement and accurate theoretical estimates of heat transport properties of mantle minerals, precluding good constrains of LM's $\kappa$[2,5,6]. Thus, the understanding of thermal conduction through the LM is still incomplete[1].

$CaSiO_3$ perovskite (CaPv) constitutes 7 vol% of a pyrolitic lower mantle[7,8]. Despite its importance, no previous experimental or theoretical estimate of cubic CaPv's lattice thermal conductivity, $\kappa_{lat}$, is available owing to its strong anharmonicity. CaPv assumes a cubic structure[9–11] under LM conditions, but it is dynamically unstable below 500 K[9,12]. At low temperatures, it presents tetragonal and orthorhombic distortions[13]. Measurements of cubic CaPv's $\kappa_{lat}$ need to be carried out above ~15 GPa and ~500 K. Calculations of $\kappa_{lat}$ are also challenging. Prevailing *ab initio* approaches relying on a perturbative treatment of weak anharmonicity have been used to evaluate $\kappa_{lat}$ of MgO periclase (Pc)[14] and pure bridgmanite, i.e., $MgSiO_3$ perovskite (MgPv)[15–17], the second and first most abundant phases of the LM, respectively. These approaches are, however, invalid for cubic CaPv.

Here, we investigate the *P-T* dependence of cubic CaPv's $\kappa_{lat}$ using both *ab initio* calculations and direct experimental measurements. We use the phonon quasiparticle approach[9,18] that can address the strongly anharmonic nature of CaPv and pulsed light heating thermoreflectance to measure $\kappa_{lat}$ in a laser-heated diamond anvil cell (DAC)[19–21]. Both calculations and experiments reveal anomalously high $\kappa_{lat}$ for CaPv. We use these results and similar ones on MgPv and Pc to estimate $\kappa_{lat}$ of a pyrolitic aggregate at LM conditions.

**Results and discussion**

**Phonon group velocities and lifetimes from phonon quasiparticles.** Heat transport in a crystalline insulator is governed by phonon-phonon scattering. According to Peierls-Boltzmann theory[22,23], the lattice thermal conductivity can be calculated as

$$\kappa_{lat} = \frac{1}{3}\sum_{\mathbf{q}s} c_{\mathbf{q}s} v_{\mathbf{q}s} l_{\mathbf{q}s}, \tag{1}$$

where $c_{\mathbf{q}s}$, $v_{\mathbf{q}s}$, $l_{\mathbf{q}s} = v_{\mathbf{q}s}\tau_{\mathbf{q}s}$ and $\tau_{\mathbf{q}s}$ are phonon heat capacity, group velocity, mean free path and lifetime, respectively, of normal mode (**q**, *s*) with frequency $\omega_{\mathbf{q}s}$. Under high temperatures ($T \geq$ 1,300 K) investigated here, $c_{\mathbf{q}s}$ is simply $k_\mathrm{B}$ in the classical limit. Thus, $v_{\mathbf{q}s}$ and $\tau_{\mathbf{q}s}$ are the critical quantities needed to evaluate $\kappa_{lat}$. For weakly anharmonic systems, $v_{\mathbf{q}s}$ and $\tau_{\mathbf{q}s}$ are routinely determined using density functional perturbation theory (DFPT) and many-body perturbation theory[24–26]. However, perturbative treatments of phonon anharmonicity become invalid in the presence of strong anharmonicity. Cubic CaPv is dynamically unstable at low temperatures and is stabilized by anharmonic fluctuations at high temperatures. We tackle the strong anharmonicity problem by using the phonon quasiparticle method[18] (see Methods). The latter has been successful in addressing this type of question[9,27], notably the unstable-to-stable transition of the cubic CaPv phase at ~600 K[9,10]. In this approach, phonon anharmonicity is expressed in terms of two quantities; renormalized phonon frequencies, $\widetilde{\omega}_{\mathbf{q}s}$, and lifetimes, $\tau_{\mathbf{q}s}$. These quantities and the phonon group velocity, $v_{\mathbf{q}s} = d\widetilde{\omega}_{\mathbf{q}s}/d\mathbf{q}$, are all temperature-dependent.

*Ab initio* simulations (see Methods) combining lattice dynamics and molecular dynamics (MD) give phonon quasiparticle properties of cubic CaPv at LM conditions. Fig. 1 shows $v_{\mathbf{q}s}$ and $\tau_{\mathbf{q}s}$ versus $\widetilde{\omega}_{\mathbf{q}s}$ collected in 135-atom MD simulations at various temperatures for $\rho$ = 4.79 g/cm³. As seen, $\widetilde{\omega}_{\mathbf{q}s}$ and $v_{\mathbf{q}s}$ have a mild and nonmonotonic temperature-dependence (Fig. 1a) while this dependence is relatively stronger for $\tau_{\mathbf{q}s}$ (Fig. 1b).

**Phonon lifetimes in the thermodynamic limit ($N \to \infty$).** To obtain $\kappa_{lat}$ converged in the thermodynamic limit, the summation in Eq. (1) must be carried out over a dense **q**-mesh throughout the Brillouin zone (BZ). The **q**-vectors sampled in *ab initio* MD simulations are limited in number by the supercell size. Therefore, it is desirable to find a parameterization of $\tau_{\mathbf{q}s}$'s dependence on frequency[28,29]. We rely on such parametrization (see Methods) to obtain $\tau_{\mathbf{q}s}$ on a dense, converged **q**-mesh. The parameterized values of $\tau_{\mathbf{q}s}$ at 1,300 K are shown in Fig. 2a. In contrast to $\tau_{\mathbf{q}s}$, $v_{\mathbf{q}s}$ can be calculated on any desired **q**-mesh, as shown in Fig. 2b. For comparison, we show $\tau_{\mathbf{q}s}$ and $v_{\mathbf{q}s}$ of MgPv[30] (see Methods) in Fig. 2c and 2d.

We have also computed the average lifetimes, $\bar{\tau}$, and average velocities, $\bar{v}$. Fig. 3a and 3b show the pressure- and temperature-dependences of CaPv and MgPv. Despite being strongly anharmonic, CaPv's $\bar{\tau}$ are longer than those of MgPv, and CaPv's $\bar{v}$ are higher at all temperatures and pressures considered. Such behavior is unexpected because MgPv is weakly anharmonic[17,18,30] and less dense than CaPv. Note that only a relatively small fraction of modes is strongly anharmonic in CaPv. Except for acoustic branches in the narrow region in the BZ marked by the dashed rectangle in Fig. 3c, most mode frequencies are weakly influenced by the strong anharmonicity. Comparing with phonon dispersions shown in Fig. 3d, the temperature-dependence of the optical modes in CaPv (5 atoms/cell) and MgPv (20 atoms/cell) are similar. Plotting these phonon dispersions in their respective BZ's highlights another important fact; cubic ($Pm\bar{3}m$) CaPv has higher symmetry and less "Bragg reflections" than orthorhombic MgPv ($Pbnm$). MgPv's phonon dispersions are folded into a smaller BZ and show less dispersive optical branches with lower velocities near the BZ edges. Branch folding in some cases causes repulsive branch interaction that reduces $v_{\mathbf{q}s}$. These branch folding effects are quite visible when comparing phonon velocities in Fig. 2b and 2d. Above ~500 cm$^{-1}$, CaPv's $v_{\mathbf{q}s}$ are generally larger than those of MgPv. Therefore, CaPv's higher structural symmetry leads to higher average quasiparticle velocities.

**Lattice thermal conductivity of CaPv.** Employing the $v_{\mathbf{q}s}$ and $\tau_{\mathbf{q}s}$ computed in the thermodynamic limit, we have calculated $\kappa_{lat}$ of CaPv at several densities ($\rho$) and temperatures using Eq. (1) (Fig. 4a). The density and temperature effects on $\kappa_{lat}$ can be described by the relation[31,32]

$$\kappa_{lat} = \kappa_{ref} \left(\frac{T_{ref}}{T}\right)^a \left(\frac{\rho}{\rho_{ref}}\right)^g, \quad (2)$$

where $g$ is[32]

$$g = b \ ln\left(\frac{\rho}{\rho_{ref}}\right) + c. \quad (3)$$

By choosing the reference density ($\rho_{ref}$) to be 4.35 g/cm³ and reference temperature ($T_{ref}$) as 1,300 K, we obtained the fitting parameters $\kappa_{ref}$, $a$, $b$ and $c$ as 10.9 W/m/K, 1.11, -10.2 and 7.87, respectively (solid lines in Fig. 4a). The choice of $\rho_{ref}$ and $T_{ref}$ is not unique but they do not change the density- and temperature-dependences of $\kappa_{lat}$. At each density, $\kappa_{lat}$ varies approximately as $1/T^{1.11}$, i.e. more abruptly than the theoretically expected ~$1/T$ relation. However, such dependence only applies within the temperature range investigated in this study, where the cubic phase is adopted, and phonon quasiparticles are well-defined.

To compare these calculations with experimental measurements or to understand the geophysical significance of these results, it is necessary to convert $\kappa_{lat}(T,\rho)$ into $\kappa_{lat}(T,P)$. We accomplish this by using an accurate thermal equation of state (EoS) of cubic CaPv. We developed such EoS by carefully combining[33,34] anharmonic $F(T,V)$ and $P(T,V)$[18,27] with experimental data[11,35] (see Methods). This procedure reduces the theoretical error in $P(T,\rho)$ caused by the exchange-correlation energy in our *ab initio* LDA calculations.

Converted and corrected results for $\kappa_{lat}(T,P)$ are shown in Fig. 4b. There are two notable features in $\kappa_{lat}(T,P)$'s behavior. First, $\kappa_{lat}$ is considerably large against the conventional wisdom that a strongly anharmonic system usually has low thermal conductivity. At the onset of the D" region at 2,600 km depth ($T = 2,735$ K and $P = 120$ GPa), $\kappa_{lat} = 16.3$ W/m/K, which is about three times as great as that of MgPv (6.5 W/m/K) (see Methods and Supplementary Fig. 1). Such roughly threefold relation applies to the entire LM. This relatively large $\kappa_{lat}$ of CaPv compared with that of MgPv results from CaPv's high structural symmetry (see discussion above and Fig. 3a and 3b). Second, $\kappa_{lat}$ varies linearly with pressure, typical behavior for minerals such as Pc and MgPv, as revealed by experiments[6,36]. However, this feature has seldomly[17,30] been reproduced in previous *ab initio* studies[14–16], reflecting the predictive power of the present approach.

Such counterintuitive results call for experimental validation, and we have measured CaPv's $\kappa_{lat}$ at high *P-T* (see Methods). The CaPv sample was prepared from pure natural wollastonite powder (Supplementary Fig. 3a) compressed in a DAC to 50 GPa at 300 K. The sample was then heated using a continuous fiber laser to more than 1,500 K. The cubic structure was confirmed by means of synchrotron X-ray diffraction (XRD) experiments at BL10XU, SPring-8. The obtained XRD patterns showed sharp peaks from cubic CaPv at high temperatures and tetragonal CaPv at 300 K after temperature quench (Supplementary Fig. 3b). Finally, to obtain $\kappa_{lat}$, a combination of pulsed light heating thermoreflectance and laser-heated DAC techniques was employed for the measurements of thermal diffusivity *in-situ* at high *P-T* conditions[21]. The measured values are summarized in Table 1 and shown in Fig. 4b. Four separated measurements were performed at ~1,300 K, and one at ~2,000 K. The measured $\kappa_{lat}$ compare well with the calculated values within experimental and computational uncertainties. Such an agreement further validates the present theoretical approach.

**Total thermal conductivity of the lower mantle.** The recently reported values of $\kappa_{lat}$ and total thermal conductivity ($\kappa_{tot}$) with a radiative contribution ($\kappa_{rad}$) considered the LM as a mixture of MgPv and Pc[15,16,31,37–39] only. In the LM both MgPv and Pc contain significant amounts of iron and MgPv also contains aluminum; therefore, it is necessary to consider the impact of these substitutions on $\kappa_{lat}$. Such an impact was shown to be substantial, reducing $\kappa_{lat}$ by ~50% for MgPv and Pc[31]. By including the impurity effects and the radiative contribution[2,16], theoretical and experimental estimations of $\kappa_{lat}$[31,37–40] and $\kappa_{tot}$[15,16] of the LM has yielded large uncertainties (Fig. 5). However, the contribution of CaPv has been omitted[15,16,31,37–39] in these studies.

With a large $\kappa_{lat}$, CaPv should contribute significantly to the $\kappa_{lat}$ of the LM. CaPv contains insignificant amounts of iron and aluminum and their effect on $\kappa_{lat}$ should be negligible. Indeed, the aluminum impurity effect is insignificant even in MgPv, especially at high temperatures[41]. Therefore, it is reasonable to approximate the *ab initio* value of CaPv's $\kappa_{lat}$ as its LM value. Our predicted $\kappa_{lat}$ along a typical LM geotherm[42] with a thermal boundary layer above the CMB is shown in Fig. 5. We compute the $\kappa_{tot}$ of the LM in the following four steps: i) $\kappa_{lat}$ of pure MgPv and pure Pc have also been investigated (see Methods) using the same phonon quasiparticle approach (see Supplementary Figs. 1 and 2, respectively); ii) as in previous studies, we assume $\kappa_{lat}$ values of MgPv and Pc are reduced by 50% after considering impurity effects[31]; iii) $\kappa_{lat}$ of a pyrolitic LM, i.e. 7 vol% CaPv, 75 vol% bridgmanite, and 18 vol% ferropericlase[7], is obtained using the Voigt-Reuss-Hill averaging[37,43] scheme (see Methods); iv) $\kappa_{tot}$ is obtained by adding the experimentally determined $\kappa_{rad}$ of a pyrolitic aggregate[44] (see Methods) to $\kappa_{lat}$. We conclude that CaPv increases $\kappa_{lat}$ of the LM by ~11% and $\kappa_{tot}$ by ~9% at all depths. $\kappa_{tot}$ of the LM along the geotherm is shown in Fig. 5, along with several other previous estimates[15,16,31,37–40]. At 2,600 km depth, we predict $\kappa_{tot}$ = 6.0(4) W/m/K. Supplementary Note 1 presents a discussion on discrepancies between different published results.

There are still large uncertainties in the estimation of $\kappa_{lat}$ of the LM mainly because of the large effect of impurities on LM phases. Relatively small variations of the Mg/Si ratio should also impact this value. However, using the common assumption that impurities reduce $\kappa_{lat}$ of MgPv and Pc by ~50%[15,31,37], CaPv should increase the thermal conductivity of the LM by ~10%. CaPv is more abundant in subducted basaltic crust[45]. In regions above the CMB populated with such basaltic materials[12], CaPv will increase the thermal conductivity significantly. Thermal conductivities of other phases expected in the lowermost mantle[45], i.e., post-perovskite, the CF phase and α-PbO$_2$-type SiO$_2$, also need to be estimated consistently before the $\kappa_{lat}$ gradient above the CMB and the heat flux across this boundary can be constrained with greater confidence.

## Methods

**Phonon quasiparticle approach.** Phonon quasiparticle properties are obtained by computing first the mode-projected velocity autocorrelation function (VAF),

$$\langle V_{\mathbf{q}s}(0) \cdot V_{\mathbf{q}s}(t) \rangle = \lim_{\tau \to \infty} \frac{1}{\tau} \int_0^\tau V_{\mathbf{q}s}^*(t') V_{\mathbf{q}s}(t' + t) dt', \quad (4)$$

where

$$V_{\mathbf{q}s}(t) = \sum_{i=1}^N V(t) \cdot e^{i\mathbf{q} \cdot \mathbf{r}_i} \cdot \hat{\mathbf{e}}_{\mathbf{q}s} \quad (5)$$

is the (**q**, *s*)-mode-projected velocity. $\hat{\mathbf{e}}_{\mathbf{q}s}$ is the harmonic phonon polarization vector of mode (**q**, *s*).

$$V(t) = V\left(\sqrt{M_1}\mathbf{v}_1(t), \ldots, \sqrt{M_N}\mathbf{v}_N(t)\right) \tag{6}$$

is the weighted velocity with $3N$ component, and $\mathbf{v}_i(t)(i = 1, \ldots, N)$ are atomic velocities calculated from MD trajectories of an $N$-atom supercell. $M_i$ is the atomic mass of the $i^{th}$ atom in the supercell. For a well-defined phonon quasiparticle, its power spectrum,

$$G_{\mathbf{q}s}(\omega) = \left|\int_0^\infty \langle V_{\mathbf{q}s}(0) \cdot V_{\mathbf{q}s}(t)\rangle e^{i\omega t} dt\right|^2 \tag{7}$$

should have a Lorentzian line shape with a peak at $\widetilde{\omega}_{\mathbf{q}s}$ and a linewidth of $1/(2\tau_{\mathbf{q}s})$. In principle, a complete decay of $\langle V_{\mathbf{q}s}(0) \cdot V_{\mathbf{q}s}(t)\rangle$ is required to obtain a reliable $G_{\mathbf{q}s}$. However, very long MD runs are needed to satisfy this condition for phonons with long lifetimes, which is inconvenient for *ab initio* MD simulations. Here we extract $\widetilde{\omega}_{\mathbf{q}s}$ and $\tau_{\mathbf{q}s}$ from $\langle V_{\mathbf{q}s}(0) \cdot V_{\mathbf{q}s}(t)\rangle$ phenomenologically from relatively shorter MD runs. For a well-defined phonon quasiparticle of mode (**q**, *s*), one can simply fit $\langle V_{\mathbf{q}s}(0) \cdot V_{\mathbf{q}s}(t)\rangle$ to the expression[46],

$$A_{\mathbf{q}s}\cos(\widetilde{\omega}_{\mathbf{q}s}t)e^{-t/(2\tau_{\mathbf{q}s})}, \tag{8}$$

where $A_{\mathbf{q}s}$ is the oscillation amplitude. In practice, to obtain reliable $\widetilde{\omega}_{\mathbf{q}s}$ and $\tau_{\mathbf{q}s}$, the fitting needs only the numerical data of $\langle V_{\mathbf{q}s}(0) \cdot V_{\mathbf{q}s}(t)\rangle$ for the first few oscillation periods[18,47].

**First-principles simulations.** We carried out *ab initio* MD simulations in the *NVT* ensemble and phonon calculations with the DFT-based Vienna *ab initio* simulation package (VASP)[24,25] employing the local density approximation (LDA) and the projected-augmented wave method (PAW)[48]. The kinetic energy cutoff adopted was 550 eV. MD simulations were conducted on $2 \times 2 \times 2$ (40 atoms) and $3 \times 3 \times 3$ supercells (135 atoms) for a series of volumes (44.39, 40.26, 36.77, 34.34 and 32.49 Å$^3$/primitive cell) corresponding to densities 4.35, 4.79, 5.25, 5.62 and 5.94 g/cm$^3$, respectively. A previous study[9] has shown that $2 \times 2 \times 2$ supercells (40 atoms) are sufficient to converge the anharmonic interaction and anharmonic phonon dispersions. In this study, we used the 3 × 3 × 3 **q**-mesh to investigate the behavior of phonon lifetimes. We carried out isochoric MD simulations for temperatures ranging from 1,300 to 4,000 K controlled by the Nosé thermostat[49] for over 60 ps with a time step of 1 fs. Throughout the volume and temperature range considered, the cubic CaPv phase was confirmed to be stable and phonon quasiparticles were well defined. Harmonic phonon frequencies and normal modes were calculated using DFPT[50] implemented in the VASP package. Anharmonic phonon dispersions and phonon lifetimes were extracted from the phonon quasiparticles sampled by the MD simulations[9,47]. Similar MD simulations were also conducted for Pc with 4 × 4 × 4 supercells (128 atoms) and *Pbnm* MgPv with 2 × 2 × 2 supercells (160 atoms). Since the focus of this paper is on CaPv, we report only a few results on these systems as Supplementary Information.

**Frequency dependence of phonon lifetimes.** Because of finite supercell sizes, $\tau_{\mathbf{q}s}$ were obtained only at relatively few discrete **q** vectors. To obtain $\tau_{\mathbf{q}s}$ at arbitrary **q**, we rely on the dependence of $\tau_{\mathbf{q}s}$ on $\widetilde{\omega}_{\mathbf{q}s}$,

$$\frac{1}{2\tau_{\mathbf{q}s}} = A\, \widetilde{\omega}_{\mathbf{q}s}^3 \qquad (9)$$

for low-frequency modes, and

$$\frac{1}{2\tau_{\mathbf{q}s}} = A\, \widetilde{\omega}_{\mathbf{q}s}^3 + B \qquad (10)$$

for high-frequency modes. *A* and *B* are parameters fitted to the *ab initio* results. We first demonstrate the effectiveness of such relations for MgPv. For the anharmonic phonon dispersion at $\rho$ = 4.21 g/cm³ and *T* = 300 K shown in Supplementary Fig. 4, we fitted Eq. (9) for phonon modes with $\widetilde{\omega}_{\mathbf{q}s}$ ≤ 192 cm⁻¹ and Eq. (10) for phonon modes with $\widetilde{\omega}_{\mathbf{q}s}$ > 192 cm⁻¹. The fitted results are shown in Supplementary Fig. 5. With these relations, we plot the distribution of $\tau_{\mathbf{q}s}$ and $v_{\mathbf{q}s}$ on an $8 \times 8 \times 8$ **q**-mesh in Supplementary Fig. 6, which is in good agreement with results obtained by the perturbative approach using the same **q**-mesh[17]. Furthermore, the resulting $\kappa_{lat}$ = 12.9 W/m/K compares well the prediction by the perturbation theory calculation, 12.5 W/m/K[17]. Note that the choice of $\widetilde{\omega}_{\mathbf{q}s}$ to split low- and high-frequency modes is not unique. Generally, the choice is guided by the goodness of the fitting. In the case of CaPv and MgPv, the highest acoustic mode frequency offered a simple criterion easy to track at different volumes and temperatures. But for Pc, the choice was solely guided by the goodness of the fitting.

We then applied the same procedure to CaPv. Supplementary Fig. 7 displays the anharmonic phonon dispersion at $\rho$ = 4.79 g/cm³ and $T$ = 1,300 K. We used Eq. (9) for phonon modes with $\widetilde{\omega}_{\mathbf{q}s} \leq 380$ cm⁻¹ and Eq. (10) for phonon modes with $\widetilde{\omega}_{\mathbf{q}s} > 380$ cm⁻¹. The fitting of calculated results to these relations is shown in Supplementary Fig. 8. By using such frequency dependences of phonon lifetimes and the phonon dispersions, we obtained the distribution of $\tau_{\mathbf{q}s}$ and $v_{\mathbf{q}s}$ on a $20 \times 20 \times 20$ **q**-mesh shown in Fig. 2a and 2b. This **q**-mesh produces converged results corresponding to the thermodynamic limit.

Similarly, using the anharmonic phonon dispersion of Pc at $\rho$ = 5.05 g/cm³ and $T$ = 2,735 K displayed in Supplementary Fig. 9, the lifetime-frequency relations for Pc were obtained by applying Eq. (9) for modes with $\widetilde{\omega}_{\mathbf{q}s} \leq 453$ cm⁻¹ and Eq. (10) for phonon modes with $\widetilde{\omega}_{\mathbf{q}s} > 453$ cm⁻¹. The fitted results are shown in Supplementary Fig. 10. The $\tau_{\mathbf{q}s}$ and $v_{\mathbf{q}s}$ interpolated on a $16 \times 16 \times 16$ **q**-mesh are summarized in Supplementary Fig. 11. The more detailed and in-depth investigation of Pc's $\kappa_{lat}$ will be discussed in another publication, but our results are very similar to those previously reported by Tang et al.[14].

**Thermal equation of state.** The thermal EoS used to convert $\kappa_{lat}(T,\rho)$ to $\kappa_{lat}(T,P)$ was obtained by fitting the calculated free energy[27],

$$F(V,T) = E(V,T_0) - T_0 S(V,T_0) - \int_{T_0}^{T} S(T')dT', \qquad (11)$$

to a third-order finite Eulerian strain expansion at each temperature. The reference temperature is $T_0 = 1{,}300$ K. $E(V,T_0)$ is the time-averaged internal energy obtained from the MD simulation at $T_0$. The entropy is

$$S_{\text{vib}} = k_B \sum_{\mathbf{q}s}[(n_{\mathbf{q}s}+1)\ln(n_{\mathbf{q}s}+1) - n_{\mathbf{q}s}\ln n_{\mathbf{q}s}], \qquad (12)$$

where $n_{\mathbf{q}s} = [\exp(\hbar\widetilde{\omega}_{\mathbf{q}s}/k_B T) - 1]^{-1}$. Renormalized phonon frequencies, $\widetilde{\omega}_{\mathbf{q}s}$, at arbitrary temperatures were obtained by fitting a second-order polynomial in $T^{18}$ to $\widetilde{\omega}_{\mathbf{q}s}$ calculated at several temperatures and constant volume.

Additional correction to $F(V,T)$ is required to have the calculated EoS in full agreement with reported experimental results[11,35]. Because experiments on CaPv were conducted around $T_{ref} = 1{,}600$ K, the correction is made at this temperature, and then this correction is applied at other temperatures. In principle, this error originates in the exchange-correlation functional adopted, the LDA, and possibly also in the selected PAWs. Anharmonicity, in principle, is adequately addressed by the quasiparticle approach. The calculated compression curve was corrected by adopting the generalized Kunc-Syassen scheme (KSr)[33,34],

$$\Delta V(P) = \frac{V_0^{exp}}{V_0^{DFT}} V\left(\frac{K_0^{DFT}}{K_0^{exp}}P, K'_{exp}\right) - V_{DFT}(P), \qquad (13)$$

where $V_0^{exp}$, $K_0^{exp}$ and $K'_{exp}$ are parameters obtained from measurements at $T_{ref}$, while $V_0^{DFT}$, $K_0^{DFT}$ and $K'_{DFT}$ are parameters obtained from $F(V,T)$ at the same $T_{ref}$. $\Delta V(P)$ can then be easily inverted to give $\Delta P(V)$. In this way, the correction to $F(V,T)$ at $T_{ref}$ can be obtained as

$$\Delta F = \int \Delta P(V) dV. \qquad (14)$$

Supplementary Fig. 12 shows that the corrected EoS is in good agreement with the experimental data within experimental uncertainties.

**Synthesis of CaSiO$_3$ perovskite.** We used pure natural wollastonite (CaSiO$_3$) powder as a starting material (Supplementary Fig. 3a). We confirmed no aluminum was included in the wollastonite sample by energy dispersive spectroscopy analysis. The powder was shaped into a disk, and then we sputtered gold or platinum onto both sides of the sample disk, which played as both a laser absorber and a monitor for temperature rise in thermal conductivity measurements. High pressure was generated in a DAC using a pair of diamond anvils with 300 μm culet and a rhenium gasket that was pre-indented to a thickness of about 50 μm. We loaded the disk-shaped sample together with a single crystal sapphire plate and NaCl pressure medium into a sample chamber with about 100 μm diameter drilled at the center of the gasket. The pressure was determined from the Raman spectrum of the diamond anvil at room temperature[51].

To synthesize CaSiO$_3$ perovskite (CaPv), we compressed the starting material to ~50 GPa at 300 K and then heated it to more than 1,500 K from both sides using a pair of continuous fiber lasers for at least 60 mins. The synthesis of CaPv was confirmed by utilizing synchrotron XRD experiments at the beamline BL10XU of SPring-8 (see sharp peaks from CaPv in Supplementary Fig. 3b)[10,52]. Subsequently, we performed thermal diffusivity measurements with changing pressure and temperature conditions. We performed thermal annealing every time after the pressure was adjusted to release deviatoric stress on the sample.

**Experimental determination of the high *P-T* lattice thermal conductivity of CaPv.** $\kappa_{lat}$ is written with thermal diffusivity *D*, density *ρ*, and isobaric heat capacity *C$_P$*;

$$\kappa_{lat} = D\rho C_P. \tag{15}$$

We obtained CaPv's $\rho$ from its high-temperature EoS (see above) and $C_P$ from thermodynamic relations and reported thermoelastic parameters of CaPv[11,35]. $D$ was measured *in-situ* at high *P-T* by employing a combination of the pulsed light heating thermoreflectance and laser-heated DAC techniques. Details of the measurement system and the analytical methods of this approach have been previously described[19–21]. We performed three separate measurements up to 67 GPa and 1950 K (Table 1). Thermal pressures in the present high *P-T* experiments were estimated empirically based on earlier laser-heated DAC measurements for CaPv[53].

**Contributions to the total thermal conductivity of the lower mantle.** We assume a pyrolitic composition for the LM, i.e., 7 vol% CaPv, 75 vol% bridgmanite, and 18 vol% ferropericlase[7]. The calculated $\kappa_{lat}$ for pure CaPv, pure MgPv, and pure Pc along the geotherm are shown as orange shaded areas in Fig. 5, Supplementary Figs. 1 and 2, respectively. At 2,600 km depth, corresponding to *P* = 120 GPa and *T* = 2,735 K, our calculated values of $\kappa_{lat}$ for CaPv, MgPv, and Pc are 16.3(10), 6.5(4) and 42.2(31) W/m/K, respectively. Considering the effects of iron and aluminum substitution in bridgmanite and ferropericlase, $\kappa_{lat}$ of these minerals are reduced by 50% from pure MgPv and pure Pc values[31], which are shown as blue shaded areas in Supplementary Figs. 1 and 2, respectively. The maximum of $\kappa_{lat}$ of a multi-phase aggregate is given by the Voigt average[37],

$$\kappa_V = \sum_i f_i \kappa_i, \tag{16}$$

and the minimum is given by the Reuss average[37],

$$\kappa_R = \left(\sum_i f_i/\kappa_i\right)^{-1}, \tag{17}$$

where $\kappa_i$ and $f_i$ are the thermal conductivity and the volume fraction of each phase $i$ in the aggregate. The aggregate $\kappa_{lat}$ is obtained using the Voigt-Reuss-Hill (VRH) average[43]

$$\kappa_{VRH} = \frac{\kappa_V + \kappa_R}{2}. \tag{18}$$

We assume the volume proportions of CaPv, bridgmanite, and ferropericlase to be 7:75:18 throughout the pyrolitic LM[7]. At LM conditions, there is also a contribution from radiative conduction to $\kappa_{tot}$. Previously reported measurements of $\kappa_{rad}$ for individual phases vary considrably[54,55]. Here we adopt a most recent measurement of $\kappa_{rad}$[44] on a pyrolitic aggregate (purple line in Fig. 5). We add this $\kappa_{rad}$ to the VRH-averaged $\kappa_{lat}$ and obtain $\kappa_{tot}$ = 6.0(4) W/m/K for the LM at 2,600 km depth, and $\kappa_{tot}$ = 4.9(3) W/m/K at the CMB.

## Data availability

The data reported in this study are available as Supplementary Information and from the corresponding author.

**Acknowledgments**

This work was funded primarily by the US Department of Energy Grant DE-SC0019759 (Z.Z. and R.M.W.) and in part by the National Science Foundation (NSF) Grant EAR-1918126 (R.M.W.). This work used the Extreme Science and Engineering Discovery Environment (XSEDE), USA, which was supported by the NSF Grant ACI-1548562. Computations were performed on Stampede2, the flagship supercomputer at the Texas Advanced Computing Center (TACC), The University of Texas at Austin generously funded by the NSF through Grant ACI-1134872. XRD measurements were performed at BL10XU, SPring-8 (Proposal No. 2019A0072 and 2019B0072) with help from Y. Ohishi. D.-B.Z. was supported by the Fundamental Research Funds for the Central Universities. K. Ohta was supported by the JSPS KAKENHI Grant 19H01995. K.H. was supported by the JSPS Grant 16H06285.

## Author contributions

R.M.W. and Z.Z. designed this project. Z.Z. performed *ab initio* calculations and modeled the results. K. Onga, A.H., K. Ohta, and K.H. were involved in thermal conductivity measurements. Z.Z., D.-B.Z., and R.M.W. drafted the manuscript. All authors discussed the results and contributed to the manuscript.

## Competing interests

The authors declare no competing interests.

**Table 1** Experimental pressures (*P*), temperatures (*T*), thermal diffusivities (*D*), and lattice thermal conductivities ($\kappa_{lat}$) of CaPv.

| Run | *P* (GPa) | *T* (K) | *D* (mm²/s) | $\kappa_{lat}$ (W/m/K) |
|-----|-----------|---------|-------------|------------------------|
| 1   | 44(2)     | 1250(120) | 3.49(53)  | 20.0(31) |
|     | 44(2)     | 1260(130) | 3.14(46)  | 18.1(27) |
| 2   | 52(2)     | 1330(130) | 3.82(59)  | 22.2(34) |
|     | 52(2)     | 1370(140) | 3.52(55)  | 20.6(32) |
| 3   | 55(3)     | 300(0)    | 16.70(247)| 37.3(55) |
|     | 67(3)     | 1950(200) | 2.84(47)  | 18.3(30) |

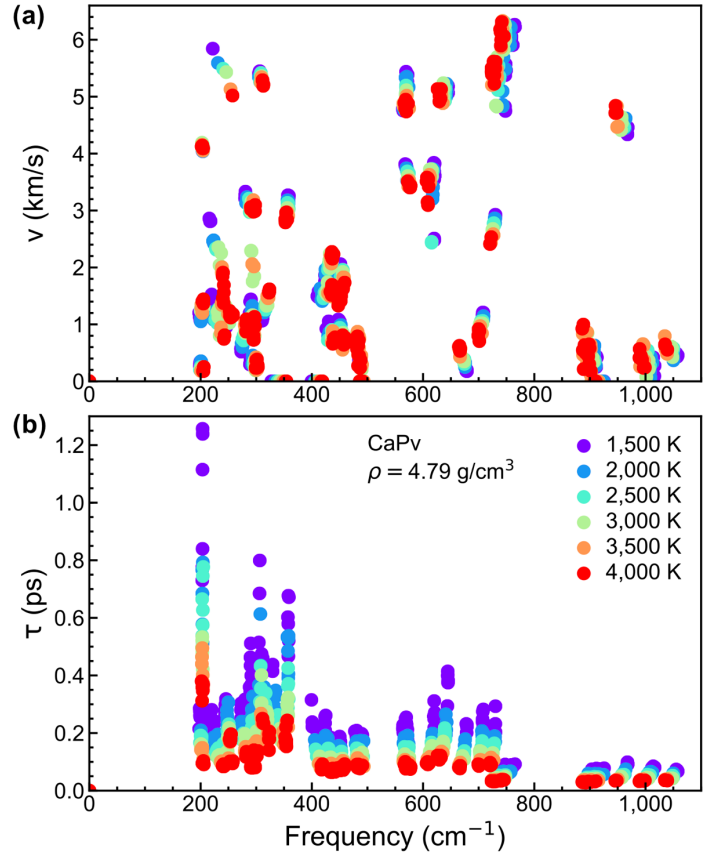

**Fig. 1 Phonon group velocities and lifetimes from phonon quasiparticles. (a)** Phonon group velocities ($v_{\mathbf{q}s}$) and **(b)** lifetimes ($\tau_{\mathbf{q}s}$) sampled by MD simulations using a $3 \times 3 \times 3$ supercell (135 atoms) vs. renormalized phonon frequency ($\widetilde{\omega}_{\mathbf{q}s}$) of CaPv at a series of temperatures.

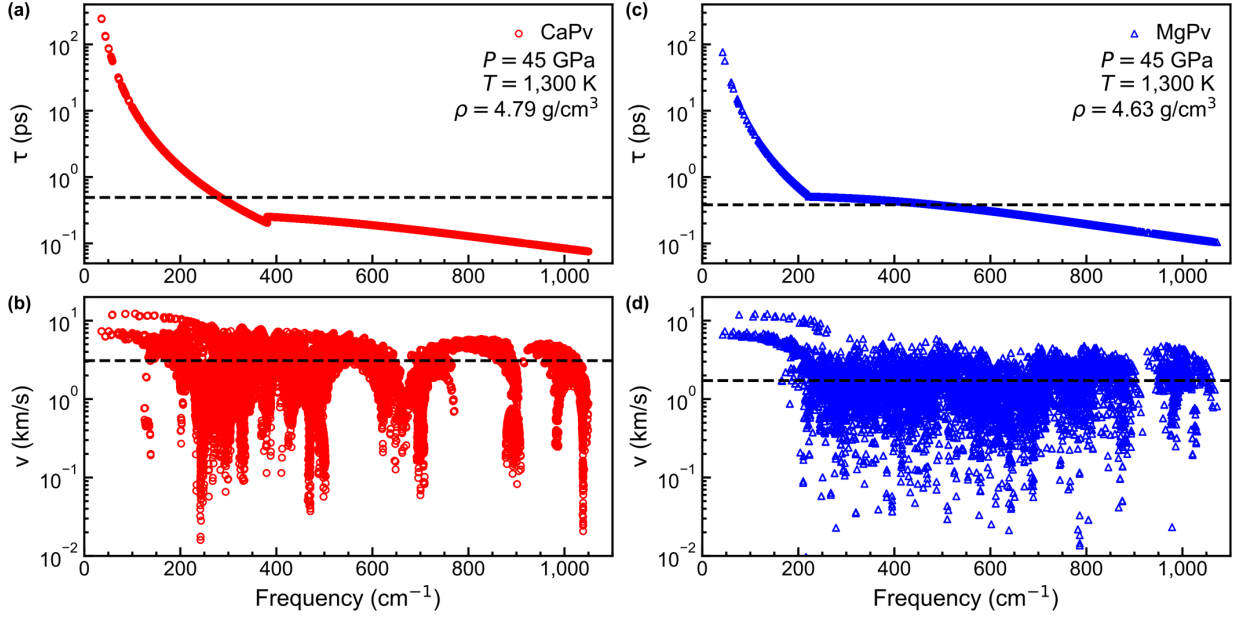

**Fig. 2 Phonon lifetimes and group velocities in the thermodynamic limit. (a)** Phonon lifetimes ($\tau_{\mathbf{q}s}$) and **(b)** group velocities ($v_{\mathbf{q}s}$) sampled by a $20 \times 20 \times 20$ **q**-mesh versus renormalized phonon frequency ($\tilde{\omega}_{\mathbf{q}s}$) of CaPv at $T = 1{,}300$ K and $\rho = 4.79$ g/cm³, corresponding to $P = 45$ GPa. **(c)** and **(d)** show $\tau_{\mathbf{q}s}$ and $v_{\mathbf{q}s}$ sampled by a $8 \times 8 \times 8$ **q**-mesh versus $\tilde{\omega}_{\mathbf{q}s}$ of MgPv at the same $P$-$T$ conditions and $\rho = 4.63$ g/cm³. Dashed lines indicate average phonon lifetimes ($\bar{\tau}$) and average phonon group velocities ($\bar{v}$).

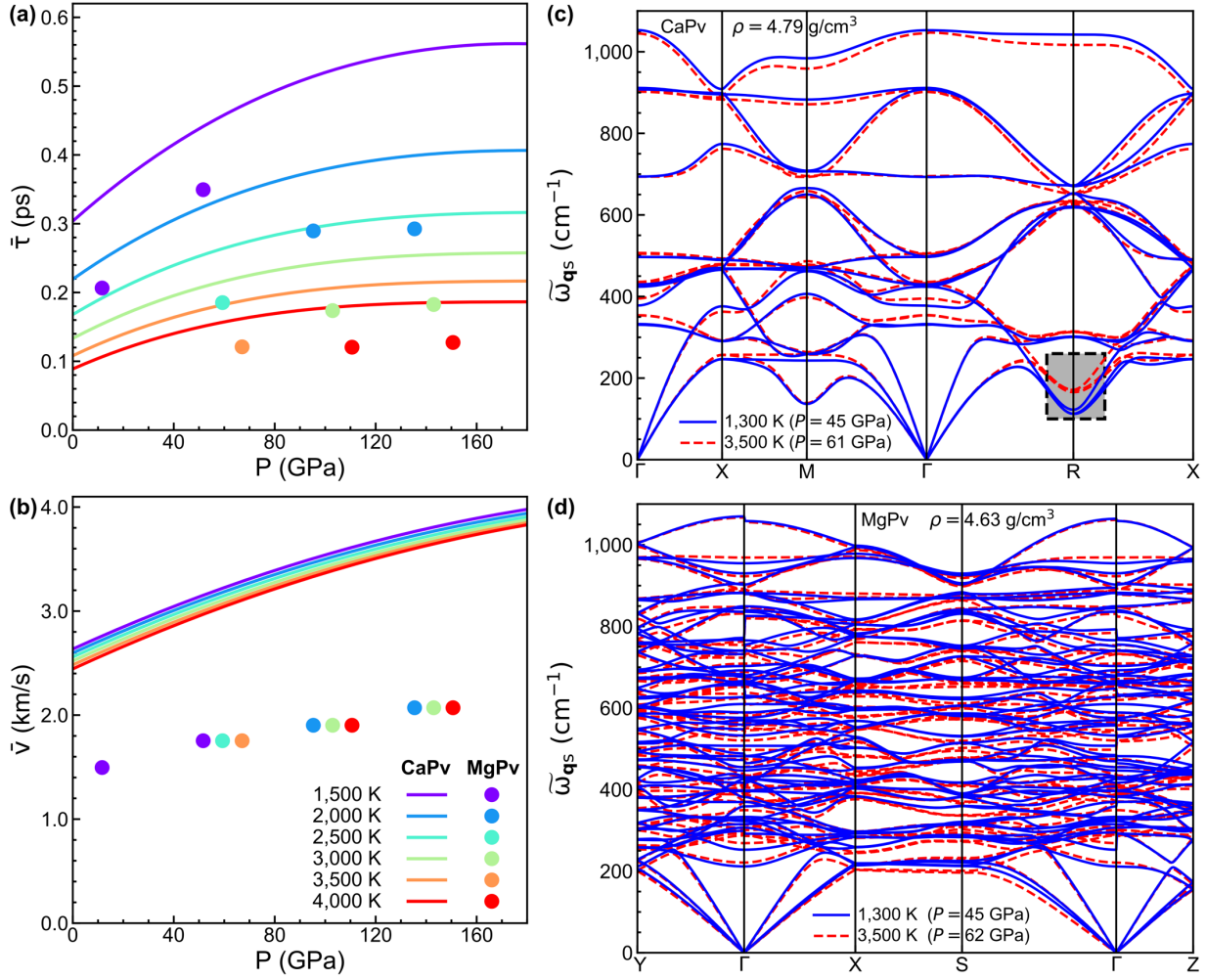

**Fig. 3 Average phonon lifetimes, group velocities, and anharmonic phonon dispersions. (a)** Average phonon lifetimes ($\bar{\tau}$) and **(b)** average phonon group velocities ($\bar{v}$) of CaPv (solid curves) and MgPv (dots) vs. $P$ at various temperatures. **(c)** Anharmonic phonon dispersions at 1,300 K (solid blue curves) and 3,500 K (dashed red curves) at $\rho = 4.79$ g/cm³ of CaPv, corresponding to $P = 45$ and 61 GPa, respectively. The dashed rectangle marks the phonon branches that are strongly temperature-dependent. **(d)** Anharmonic phonon dispersions at 1,300 K (solid blue curves) and 3,500 K (dashed red curves) at $\rho = 4.63$ g/cm³ of MgPv, corresponding to $P = 45$ and 62 GPa, respectively.

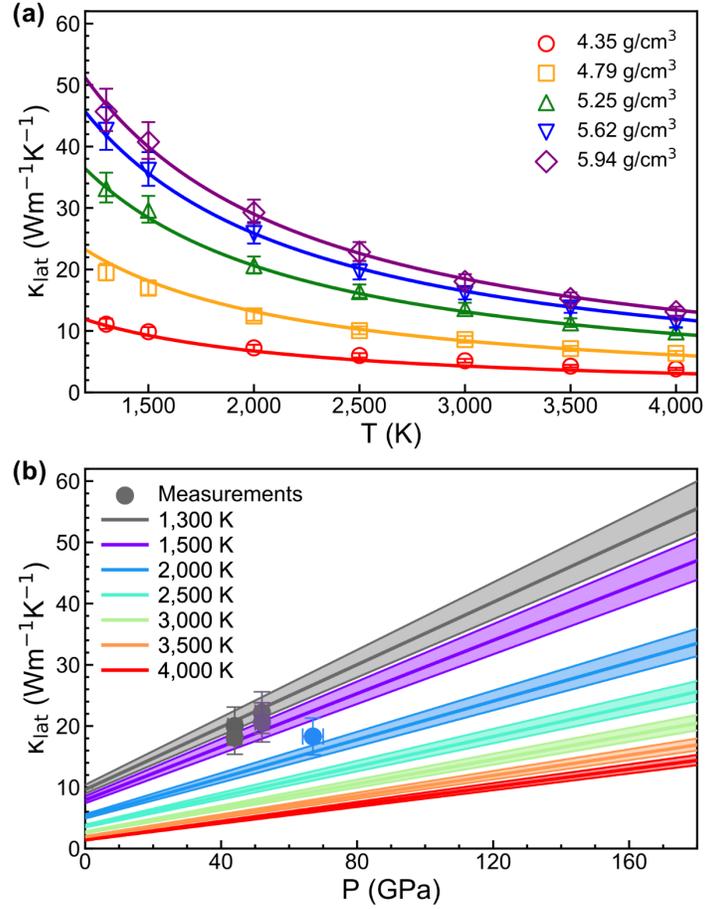

**Fig. 4 Temperature and pressure dependences of CaPv's lattice thermal conductivity. (a)** Lattice thermal conductivity ($\kappa_{lat}$) vs. $T$ of CaPv (symbols) for a series of densities. Error bars show the computational uncertainties. Solid curves show the fitting to Eq. (2). **(b)** $\kappa_{lat}$ vs. $P$ of CaPv at a series of temperatures. Shaded areas indicate computational uncertainties. Solid symbols with error bars are experimental measurements.

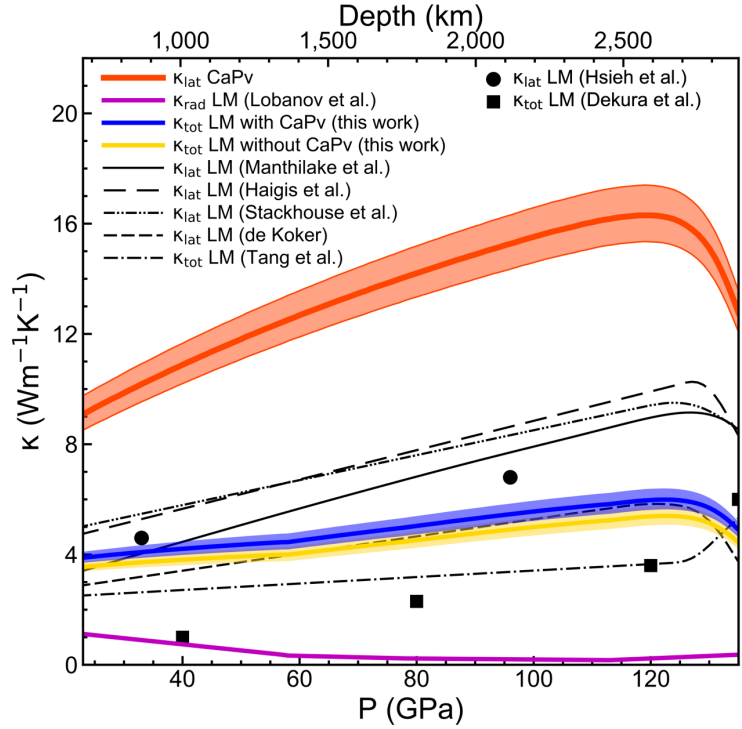

**Fig. 5 Thermal conductivities of CaPv and of a pyrolitic aggregate along the geotherm[42].** The lattice thermal conductivity ($\kappa_{lat}$) of CaPv is shown in orange, the experimental radiative thermal conductivity[44] ($\kappa_{rad}$) of a pyrolitic aggregate is shown in purple, the total thermal conductivity ($\kappa_{tot}$) of the LM without CaPv is shown in yellow and with CaPv in blue. Shaded areas indicate computational uncertainties. Available estimations of $\kappa_{lat}$[31,37–40] and $\kappa_{tot}$[15,16] of the LM are shown for comparison.

# Supplementary Information
## Thermal Conductivity of CaSiO$_3$ Perovskite at Lower Mantle Conditions
Zhang et al.

**The supplementary information consists of Supplementary Note 1, Supplementary Table 1 and 2, and Supplementary Figs. 1 to 13.**

**Supplementary Note 1**

**Estimation of the thermal conductivity of the LM.**

Previous studies[1–7] and the present one on the lattice thermal conductivity ($\kappa_{lat}$) and total thermal conductivity ($\kappa_{tot}$) of the pyrolitic lower mantle (LM) show significant uncertainties, as shown in Fig. 5 in the main text. At 2,600 km depth, corresponding to $P$ = 120 GPa and $T$ = 2,735 K, estimated thermal conductivities range from ~3 W/m/K with the inclusion of radiative thermal conductivity ($\kappa_{rad}$) to ~10 W/m/K without $\kappa_{rad}$. In general, $\kappa_{lat}$ and $\kappa_{tot}$ of the LM reported by Manthilake et al.[1], Haigis et al.[2], Stackhouse et al.[3], and Hsieh et al.[7] are larger than ours, while those reported by de Koker[4], Tang et al.[5], and Dekura et al.[6] are smaller. These studies share the common assumption that the pyrolitic LM consists of approximately 80 vol% bridgmanite and 20 vol% ferropericlase. Supplementary Table 1 summarized the LM composition adopted in different studies. Stackhouse et al. also included CaPv, but CaPv's $\kappa_{lat}$ used in this work was estimated by scaling MgPv's $\kappa_{lat}$[3]. $\kappa_{rad}$ is not included in the studies by Manthilake et al., Haigis et al., Stackhouse et al., Hsieh et al. and de Koker, while $\kappa_{rad}$ in Tang et al. and Dekura et al. does not enhance their $\kappa_{tot}$ significantly. We have adopted a recently published $\kappa_{rad}$ for the pyrolitic LM, which is relatively small, i.e., <0.5 W/m/K in most of the LM[8]. Therefore, the discrepancy between various estimates of LM's $\kappa_{tot}$ are due to different values of bridgmanite's and ferropericlase's $\kappa_{lat}$. Another significant source of discrepancy is the estimated effect of aluminum and iron substitutions in the $\kappa_{lat}$ of pure MgSiO$_3$ perovskite (MgPv) and pure MgO periclase (Pc). The $\kappa_{lat}$ of pure MgPv and of pure Pc used in this and previous studies[1–7,9–12] at 2,600 km depth are summarized in Supplementary Table 2.

To estimate the LM's $\kappa_{lat}$, Manthilake et al.[1] measured $\kappa_{lat}$ of pure, aluminum-, and iron-bearing MgPv up to 1,073 K, and of pure Pc, and ferropericlase up to 1,273 K at relatively low pressures. These $\kappa_{lat}$ were then extrapolated to the LM conditions using a thermodynamic model[1]. Due to aluminum and iron substitutions, $\kappa_{lat}$ of bridgmanite and ferropericlase are predicted to decrease by ~50% from those of pure MgPv and Pc, respectively. We also adopted a similar reduction factor in our work. $\kappa_{lat}$ for MgPv and Pc measured by Manthilake et al.[1] are compared with ours in Supplementary Fig. 13a, 13b and 13c. For MgPv, measured and calculated $\kappa_{lat}$ agree better at low temperatures. The experimental and theoretical $T$-dependence of $\kappa_{lat}$ differ with a $1/T^{0.43}$ dependence in the measurements and a $1/T^{1.02}$ dependence obtained in the present study. This difference makes the extrapolated $\kappa_{lat}$ of MgPv based on Manthilake et al.'s model considerably larger than our values at LM conditions. As for Pc, the measured $\kappa_{lat}$ is smaller than our calculated values at low temperatures. However, the measured $T$-dependence, $1/T^{0.76}$, is more gentle than ours, $1/T^{1.19}$. Therefore, their extrapolated values and our calculated values at high temperatures agree better. Supplementary Table 2 displays values of $\kappa_{lat}$ for MgPv and Pc at 2,600 km depth according to these studies. Pc's $\kappa_{lat}$ based on Manthilake et al.'s model is only ~15% smaller than ours. Such difference won't impact much on the estimation of LM's $\kappa_{lat}$ since there is less than 20 vol% of Pc in this region. However, the extrapolated MgPv's $\kappa_{lat}$ is almost three times as great as our calculated value. Since MgPv is the dominating LM phase, such difference impacts significantly the pyrolitic aggregate averaged $\kappa_{lat}$, resulting in an LM's $\kappa_{lat}$ considerably larger than ours. Hsieh et al.[7] measured $\kappa_{lat}$ of aluminum- and iron-bearing bridgmanite and iron-bearing ferropericlase to high pressures at 300 K. By assuming the $T$-dependence of $\kappa_{lat}$ of these phases to be $1/T^{0.5}$, the LM's $\kappa_{lat}$ was estimated. At 2,600 km depth, their extrapolated $\kappa_{lat}$ of bridgmanite and ferropericlase are approximately 6.9 and 13.4 W/m/K, respectively. According to Hsieh et al.'s measurements, iron substitution can reduce $\kappa_{lat}$ of Pc by a factor of ~10 at ambient conditions[7]. $\kappa_{lat}$ of pure MgPv[11] and pure Pc[12] at 2,600 km depth obtained by Hsieh and collaborators in previous works are shown in Supplementary Table 2. Their adopted effect of impurities in $\kappa_{lat}$ of MgPv and Pc are roughly a 25% and 50% reduction, respectively.

Calculated $\kappa_{lat}$ values of MgPv and Pc by Haigis et al.[2], Stackhouse et al.[3,9], de Koker[4], Tang et al.[5,10], and Dekura et al.[6] at 2,600 km depth are shown in Supplementary Table 2. $\kappa_{lat}$ of MgPv and Pc by Haigis et al.'s are roughly twice larger than ours. By assuming a 50% reduction in $\kappa_{lat}$ due to impurities, the resulting $\kappa_{lat}$ of the LM remains twice larger than ours. Stackhouse et al.'s $\kappa_{lat}$ of MgPv and Pc are comparable to ours. However, they assumed a small 8% reduction in $\kappa_{lat}$ due to impurities at core-mantle boundary conditions, resulting in a considerably larger value of LM's $\kappa_{lat}$ compared to our values. de Koker reported a $\kappa_{lat}$ for Pc comparable to ours but much smaller values for MgPv. Using the same $\kappa_{lat}$ reduction factor due to impurities, de Koker's $\kappa_{lat}$ for the LM should be considerably smaller than ours because of MgPv's dominant role. Nevertheless, de Koker reported a $\kappa_{lat}$ for the LM similar to ours. This is because de Koker assumed the $\kappa_{lat}$ reduction due to impurities to be counteracted by $\kappa_{rad}$. In other words, de Koker estimated the LM's $\kappa_{lat}$ using values of pure MgPv and pure Pc and did not explicitly include an estimation of LM's $\kappa_{rad}$. Tang et al.'s $\kappa_{lat}$ of Pc agrees very well with our calculated value, while MgPv's $\kappa_{lat}$ is much smaller than ours. As for impurity effects, Tang et al. adopted a larger iron substitution effect for Pc and much weaker effect for MgPv. At 2,600 km depth, $\kappa_{lat}$ of Pc and MgPv are reduced by roughly 75% and 25%, respectively. After including $\kappa_{rad}$[5], Tang et al. reported $\kappa_{lat}$ for the LM generally smaller than ours, mainly because of the much smaller $\kappa_{lat}$ of MgPv. Dekura et al. obtained a $\kappa_{lat}$ for MgPv lower than ours and adopted Stackhouse et al.[9]'s $\kappa_{lat}$ for Pc, which is also smaller than ours. After a 50% $\kappa_{lat}$ reduction due to impurities and adding the $\kappa_{rad}$[13], Dekura et al. obtained $\kappa_{lat}$ for the LM generally smaller than our values. The increase in $\kappa_{lat}$ in the D" region reported by Tang et al. and Dekura et al. are caused by the dominating role of their estimated $\kappa_{rad}$[5,13].

As pointed out in the main text, a significant reason for discrepancies between the *ab initio* calculations using perturbation theory (Dekura et al., Tang et al.) and the quasiparticle approach (de Koker and ours) is primarily the **q**-point sampling (# of vibrational modes). Molecular dynamics (MD), whether using an *ab initio* method (Stackhouse et al.) or classical force fields (Haigis et al.), sample fewer vibrational modes than is possible with the other approaches, even though extrapolations to the thermodynamic limit are common. Vibrational mode sampling strongly affects the calculated values of $\kappa_{lat}$ in perfect crystals. Another source of uncertainty in $\kappa_{lat}$ calculations are the interpolation of phonon lifetimes vs. frequency, $\tau(\omega)$. For example, Dekura et al. calculated $\tau$ for MgPv at the Brillouin zone center and used the empirical $1/\omega^2$ relationship to estimate $\tau$ away from the zone center. Hence, they obtained a slightly smaller value of $\kappa_{lat}$ than ours, since we used a $1/\omega^3$ dependence for $\tau$. Tang et al. calculated MgPv's $\tau$ on an $8 \times 8 \times 6$ **q**-mesh, and their $\tau$ at the zone center agree well with Dekura et al.'s results. However, Tang et al. demonstrated the failure of the $1/\omega^2$ dependence of $\tau$ away from the zone center and report much smaller values of $\kappa_{lat}$. Because pure Pc and pure MgPv are weakly anharmonic, perturbation theory and the quasiparticle method should give comparable results for good interpolations of $\tau(\omega)$ when needed. Our results for MgPv and Pc are fully converged w.r.t. to **q**-point sampling, as documented in the comparison (see main text) with another perturbation theory calculation for MgPv[14] with converged **q**-mesh and the good agreement between Tang et al.[10]'s results and ours for Pc using the same well converged **q**-mesh. These agreements also indicate that the $1/\omega^3$ dependence of $\tau$ provides a reasonably satisfactory interpolation for $\tau(\omega)$.

Our consistent and converged calculations give $\kappa_{lat}^{Pc} > \kappa_{lat}^{CaPv} > \kappa_{lat}^{MgPv}$. This result emphasizes the significant role played by CaPv in LM's $\kappa_{lat}$. The good agreement between *ab initio* predictions and experimental data on CaPv further validates the quasiparticle method used in this study. Great uncertainties remain on the effect of impurities and $\kappa_{rad}$ on the LM's $\kappa_{tot}$.

**Supplementary Table 1 Composition of the pyrolitic LM in this study and previous studies;** Manthilake et al.[1], Haigis et al.[2], Stackhouse et al.[3], de Koker[4], Tang et al.[5], Dekura et al.[6], and Hsieh et al.[7].

| proportion (vol%) | this work | Manthilake et al. | Haigis et al. | Stackhouse et al. | de Koker | Tang et al. | Dekura et al. | Hsieh et al. |
|---|---|---|---|---|---|---|---|---|
| bridgmanite | 75 | 80 | 82 | 75 | 80 | 80 | 80 | 80 |
| ferropericlase | 18 | 20 | 18 | 19 | 20 | 20 | 20 | 20 |
| CaPv | 7 | 0 | 0 | 6 | 0 | 0 | 0 | 0 |

**Supplementary Table 2 Lattice thermal conductivities ($\kappa_{lat}$) of MgPv and Pc at 2,600 km depth of the LM reported in this study and previous studies;** Manthilake et al.[1], Haigis et al.[2], Stackhouse et al.[3,9], de Koker[4], Tang et al.[5,10], Dekura et al.[6] and Hsieh et al.[7,11,12].

| $\kappa_{lat}$ (W/m/K) | this work | Manthilake et al. | Haigis et al. | Stackhouse et al. | de Koker | Tang et al. | Dekura et al. | Hsieh et al. |
|---|---|---|---|---|---|---|---|---|
| MgPv | 6.5 | 17.4 | 13.8 | 7.6 | 2.6 | 2.0 | 5.1 | 9.1 |
| Pc | 42.2 | 35.1 | 70.8 | 30.5 | 36.9 | 44.8 | 30.5 | 26.4 |

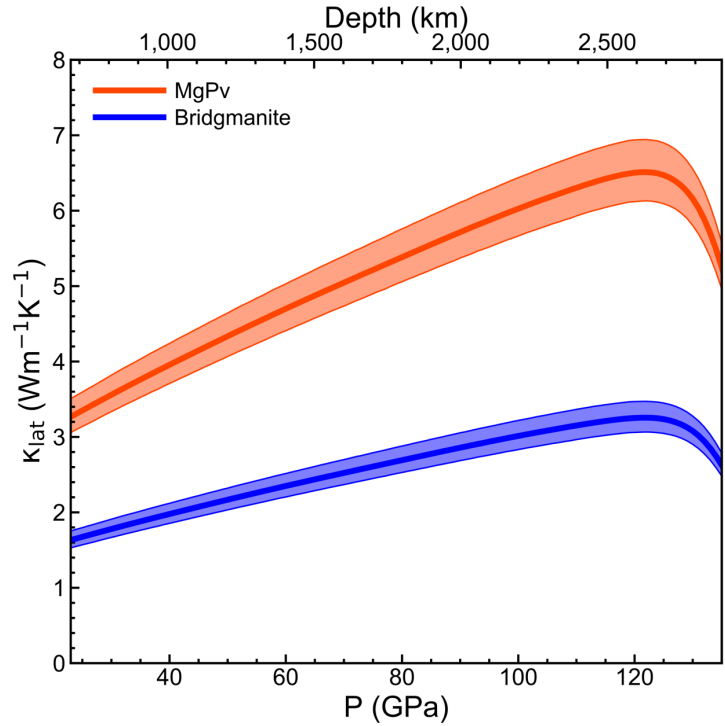

**Supplementary Figure 1 - $\kappa_{lat}$ of MgPv along the geotherm.** The $\kappa_{lat}$ of pure MgPv along the geotherm[15] is shown in orange. $\kappa_{lat}$ of bridgmanite (blue) is reduced by 50% from pure MgPv values due to iron and aluminum impurity effects[1]. Shaded areas indicate computational uncertainties.

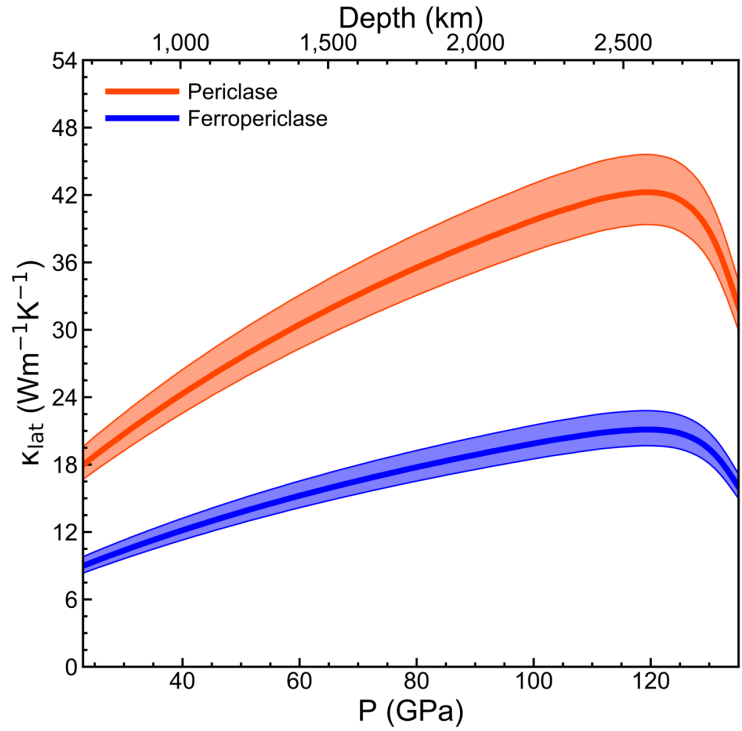

**Supplementary Figure 2 - $\kappa_{lat}$ of Pc along the geotherm.** The $\kappa_{lat}$ of pure Pc along the geotherm[15] is shown in orange. $\kappa_{lat}$ of ferropericlase (blue) is reduced by 50% from pure Pc values due to iron and aluminum impurity effects[1]. Shaded areas indicate computational uncertainties.

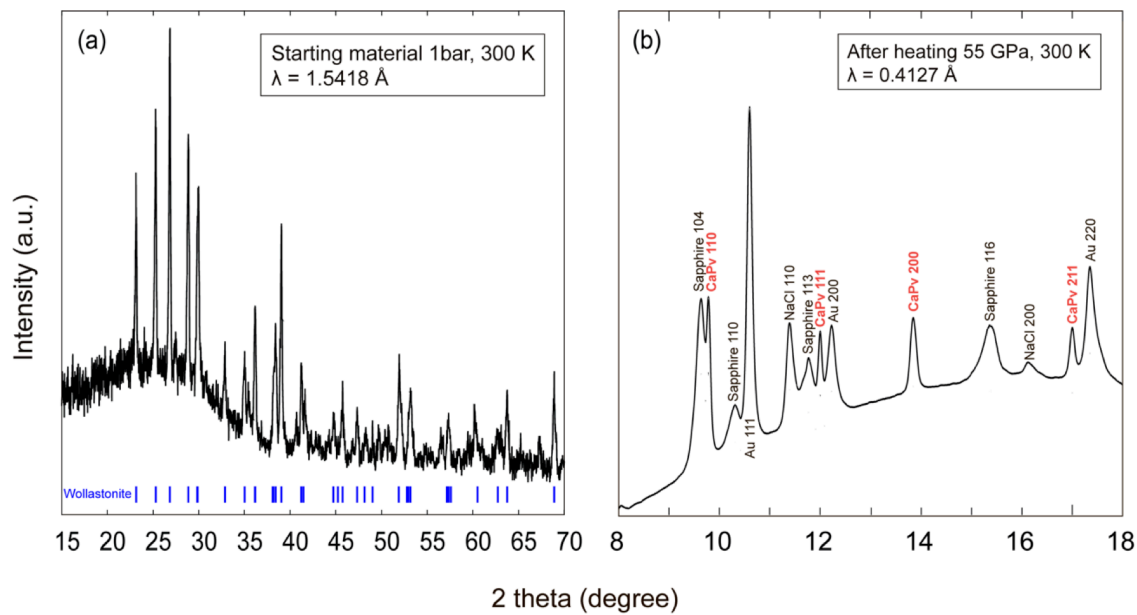

**Supplementary Figure 3 - XRD patterns of CaSiO₃.** XRD patterns of **(a)** wollastonite starting material and **(b)** synthesized CaPv after temperature quench. The calculated XRD peak positions of wollastonite are shown by blue bars. Au: gold foil, sapphire and NaCl: thermal insulator and pressure medium.

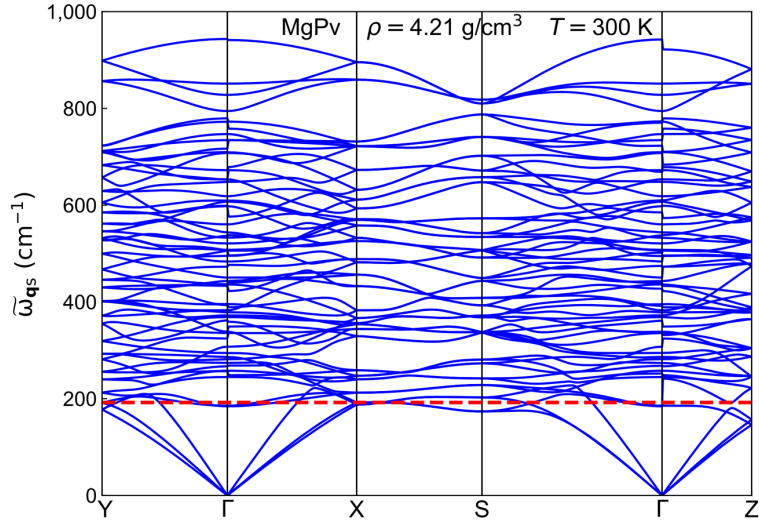

**Supplementary Figure 4 - Splitting of MgPv's low- and high-frequency modes.** Anharmonic phonon dispersions (blue solid lines) of MgPv at $\rho$ = 4.21 g/cm$^3$ and $T$ = 300 K. The red dashed line shows $\widetilde{\omega}_{\mathbf{q}s}$ = 192 cm$^{-1}$ used to split the low- and the high-frequency modes.

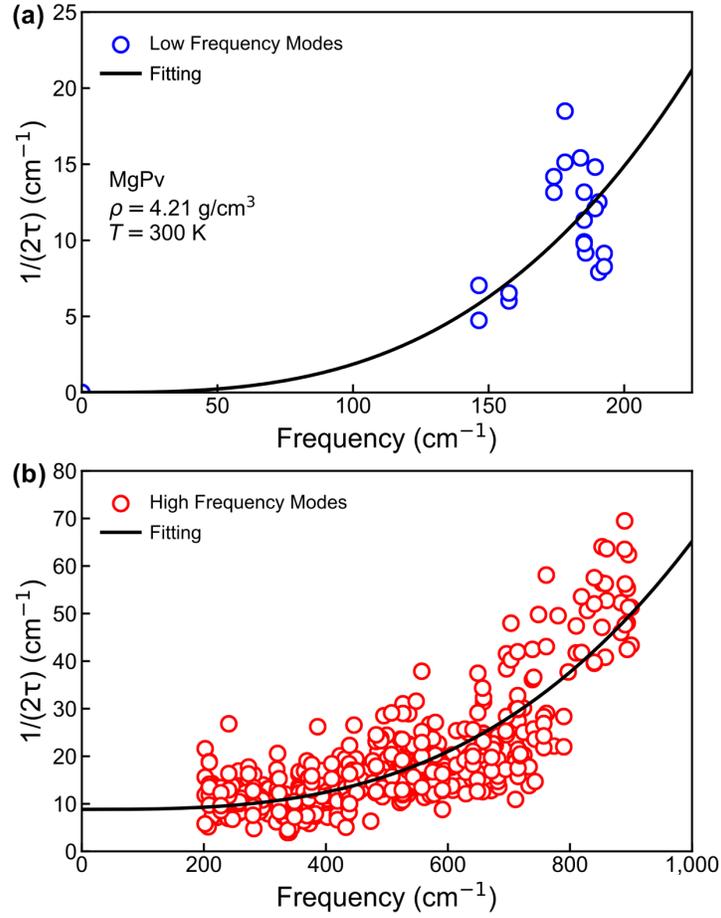

**Supplementary Figure 5 - Frequency dependence of phonon lifetimes of MgPv.** Linewidth, $1/(2\tau_{\mathbf{q}s})$, of phonon quasiparticle as a function of renormalized phonon frequency, $\widetilde{\omega}_{\mathbf{q}s}$, of MgPv at $\rho$ = 4.21 g/cm³ and $T$ = 300 K for **(a)** low-frequency modes, and **(b)** high-frequency modes sampled by 2 × 2 × 2 supercells (hollow circles) in the MD simulations. Solid lines are fitting for frequency-lifetime relation: **(a)** $1/(2\tau_{\mathbf{q}s}) = 1.9 \times 10^{-6} \widetilde{\omega}_{\mathbf{q}s}^3 /\text{cm}^{-2}$, and **(b)** $1/(2\tau_{\mathbf{q}s}) = 5.6 \times 10^{-8} \widetilde{\omega}_{\mathbf{q}s}^3 /\text{cm}^{-2} + 8.8 \text{cm}^{-1}$.

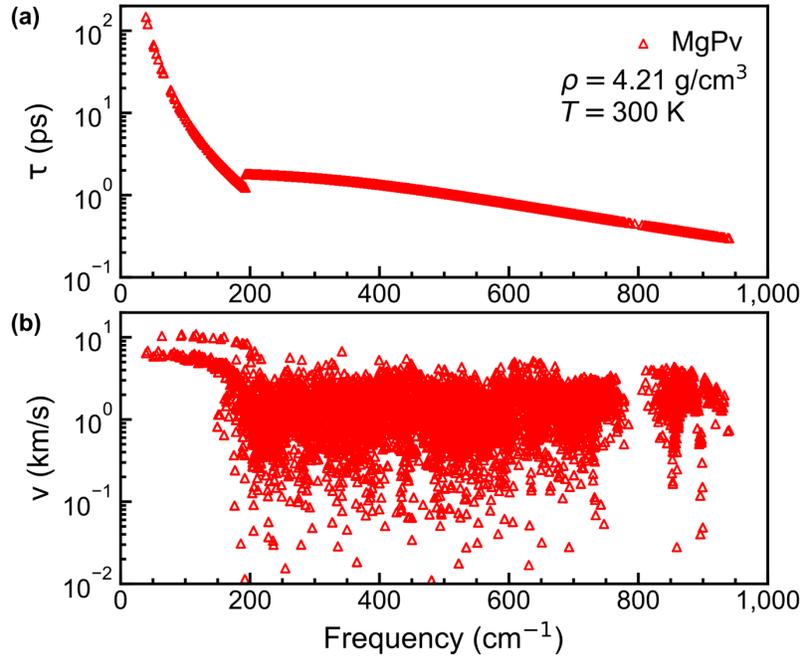

**Supplementary Figure 6 - Phonon lifetimes and group velocities of MgPv in the thermodynamic limit.** **(a)** Phonon lifetimes ($\tau_{\mathbf{q}s}$) and **(b)** group velocities ($v_{\mathbf{q}s}$) sampled by $8 \times 8 \times 8$ **q**-mesh versus renormalized phonon frequency ($\widetilde{\omega}_{\mathbf{q}s}$) of MgPv at $\rho$ = 4.21 g/cm$^3$ and $T$ = 300 K.

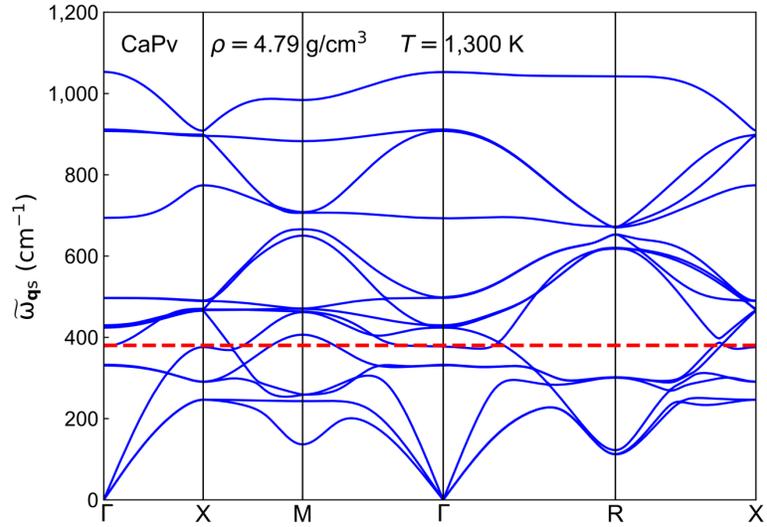

**Supplementary Figure 7 - Splitting of CaPv's low- and high-frequency modes.** Anharmonic phonon dispersions (blue solid lines) of CaPv at $\rho$ = 4.79 g/cm$^3$ and $T$ = 1,300 K. The red dashed line shows $\widetilde{\omega}_{\mathbf{q}s}$ = 380 cm$^{-1}$ was used to split the low- and the high-frequency modes.

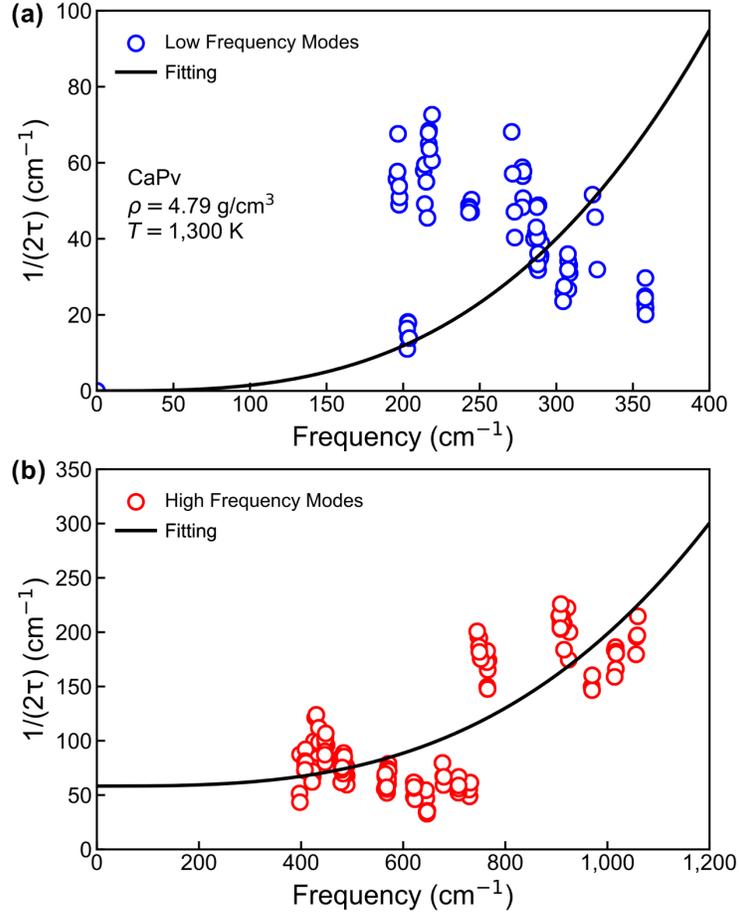

**Supplementary Figure 8 - Frequency dependence of phonon lifetimes of CaPv.** Linewidth, $1/(2\tau_{\mathbf{q}s})$, of phonon quasiparticles as a function of renormalized phonon frequency, $\widetilde{\omega}_{\mathbf{q}s}$, of CaPv at $\rho$ = 4.79 g/cm³ and $T$ = 1,300 K for **(a)** low-frequency modes, and **(b)** high-frequency modes sampled by 3 × 3 × 3 supercells (hollow circles) in the MD simulations. Solid lines are fitting for frequency-lifetime relation: **(a)** $1/(2\tau_{\mathbf{q}s}) = 1.5 \times 10^{-6}\widetilde{\omega}_{\mathbf{q}s}^3/\text{cm}^{-2}$, and **(b)** $1/(2\tau_{\mathbf{q}s}) = 1.4 \times 10^{-7}\widetilde{\omega}_{\mathbf{q}s}^3/\text{cm}^{-2} + 58\text{cm}^{-1}$.

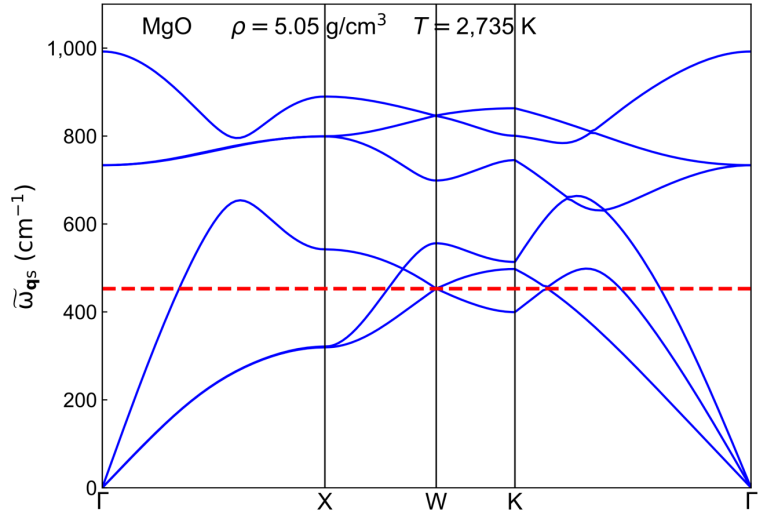

**Supplementary Figure 9 - Splitting of Pc's low- and high-frequency modes.** Anharmonic phonon dispersions (blue solid lines) of Pc at $\rho = 5.05$ g/cm$^3$ and $T = 2{,}735$ K. The red dashed line shows $\widetilde{\omega}_{\mathbf{q}s} = 453$ cm$^{-1}$ was used to split the low- and the high-frequency modes.

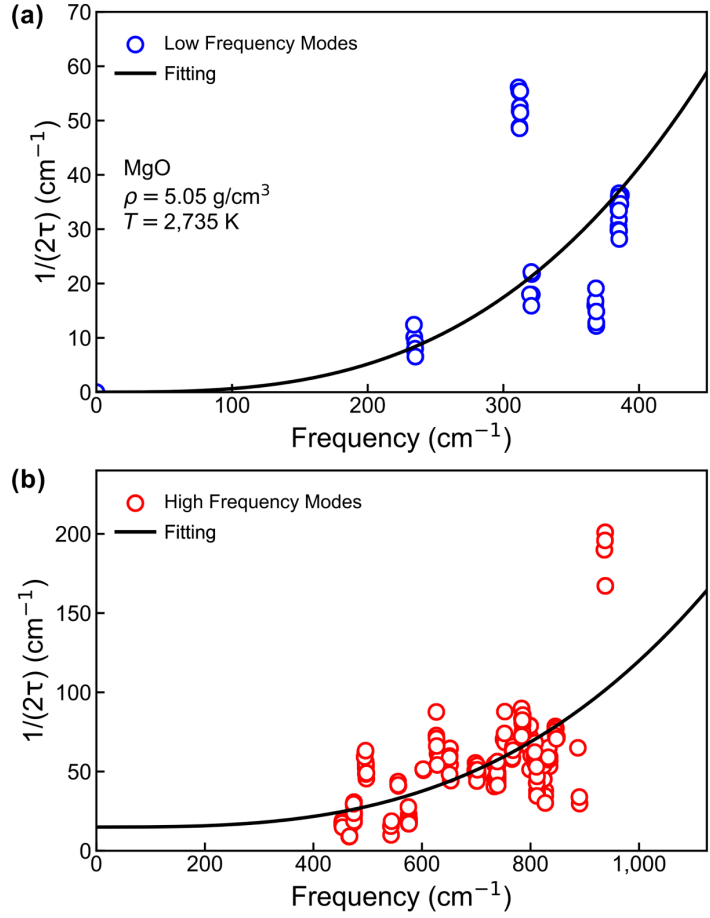

**Supplementary Figure 10 - Frequency dependence of phonon lifetimes of Pc.** Linewidth, $1/(2\tau_{\mathbf{q}s})$, of phonon quasiparticles as a function of renormalized phonon frequency, $\widetilde{\omega}_{\mathbf{q}s}$, of Pc at $\rho = 5.05$ g/cm$^3$ and $T = 2{,}735$ K for **(a)** low-frequency modes, and **(b)** high-frequency modes sampled by $4 \times 4 \times 4$ supercells (hollow circles) in the MD simulations. Solid lines are fitting for frequency-lifetime relation: **(a)** $1/(2\tau_{\mathbf{q}s}) = 6.5 \times 10^{-7} \widetilde{\omega}_{\mathbf{q}s}^3 / \text{cm}^{-2}$, and **(b)** $1/(2\tau_{\mathbf{q}s}) = 1.0 \times 10^{-7} \widetilde{\omega}_{\mathbf{q}s}^3 / \text{cm}^{-2} + 15 \text{cm}^{-1}$.

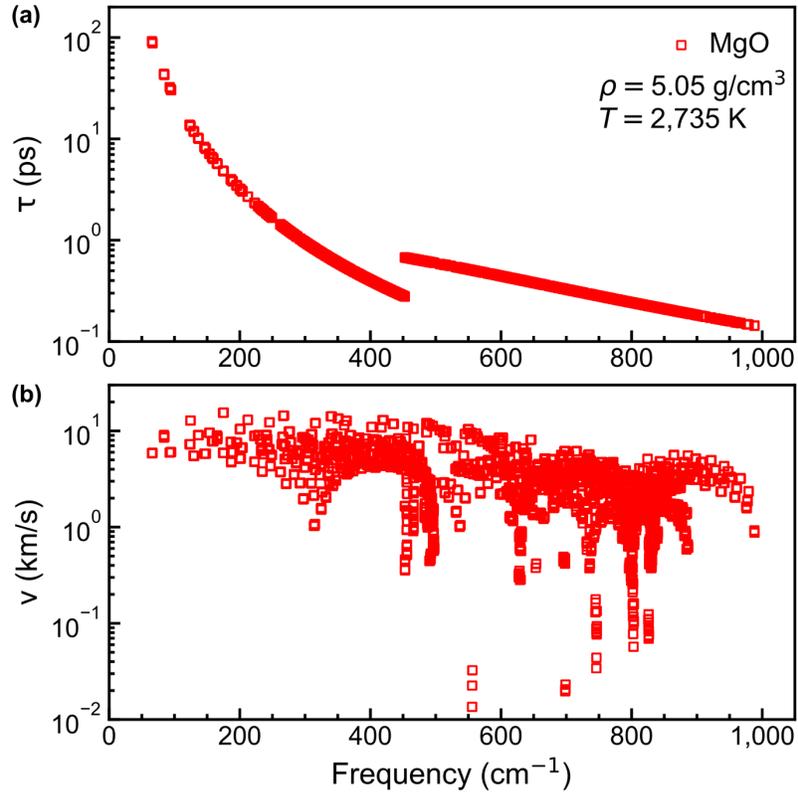

**Supplementary Figure 11 - Phonon lifetimes and group velocities of Pc in the thermodynamic limit. (a)** Phonon lifetimes ($\tau_{\mathbf{q}s}$) and **(b)** group velocities ($v_{\mathbf{q}s}$) sampled by $16 \times 16 \times 16$ **q**-mesh versus renormalized phonon frequency ($\widetilde{\omega}_{\mathbf{q}s}$) of MgPv at $\rho = 5.05$ g/cm³ and $T = 2{,}735$ K.

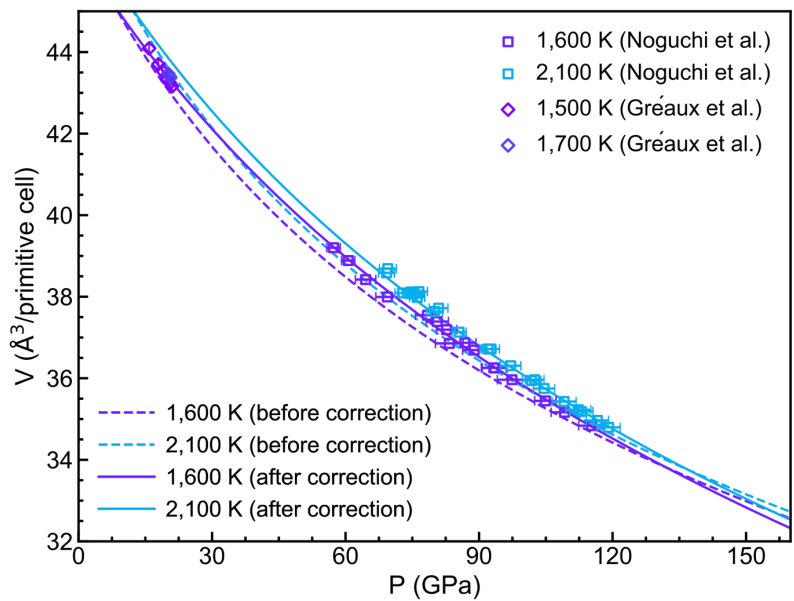

**Supplementary Figure 12 - Thermal EoS of CaPv.** Isothermal third-order finite strain equation of state (EoS) of CaPv at $T$ = 1,600 and 2,100 K before (dashed lines) and after (solid lines) DFT energy correction compared with experimental results reported by Noguchi et al.[16] (squares) and Gréaux et al.[17] (diamonds). Error bars show the experimental uncertainties.

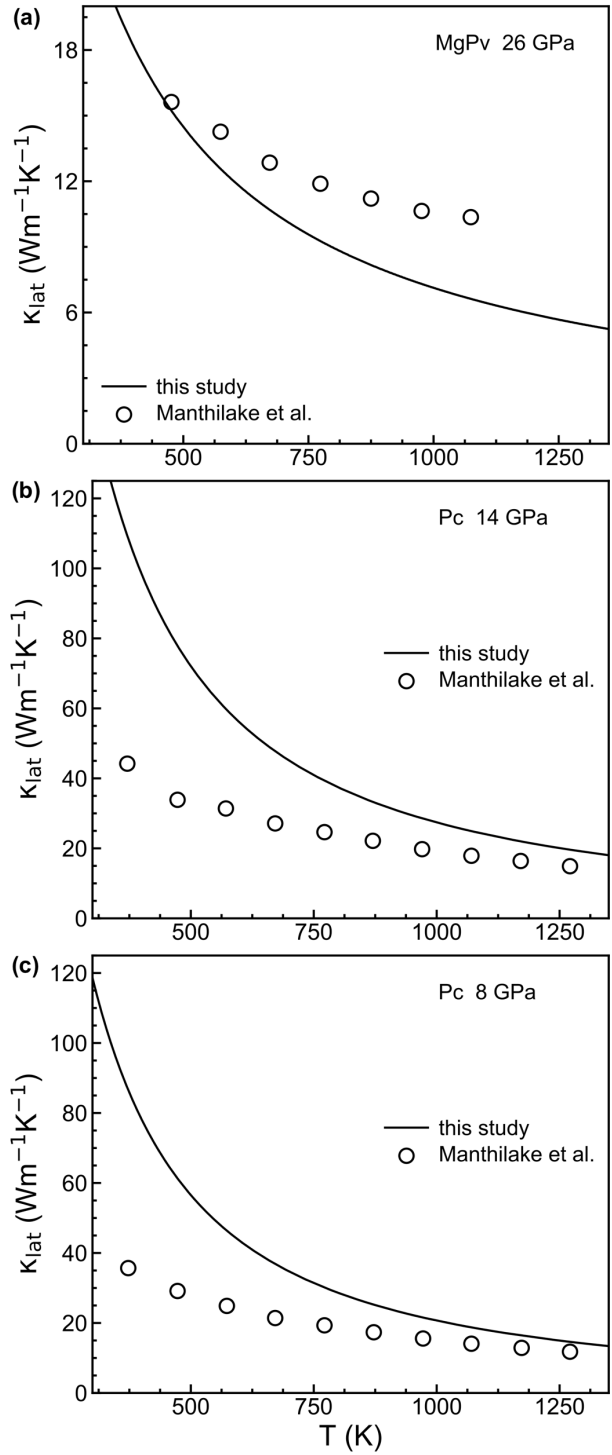

**Supplementary Figure 13 - $\kappa_{lat}$ of MgPv and Pc.** Comparison of $\kappa_{lat}$ calculated in this study and measured by Manthilake et al.[1] for **(a)** MgPv, **(b)** and **(c)** Pc.